\documentclass[aps,prl,twocolumn,showpacs]{revtex4}  

\usepackage{blindtext}
\usepackage{graphicx}
\usepackage{dcolumn}
\usepackage{bm}
\usepackage{amssymb}

\begin{document}

\title{Quasinormal modes of BTZ black hole and Hawking-like radiation in
graphene}
\author{B. S. Kandemir}
\thanks{Corresponding author}
\email{kandemir@science.ankara.edu.tr}
\author{\"Umit Ertem}
\email{umitertemm@gmail.com}
\affiliation{Department of Physics,
Ankara University, Faculty of Sciences, 06100, Tando\u gan-Ankara,
Turkey\\}
\date{\today}

\begin{abstract}
The Ba\~{n}ados-Teitelboim-Zanelli (BTZ) black hole model corresponds to a
solution of (2+1)-dimensional Einstein gravity with negative cosmological
constant, and by a conformal rescaling its metric can be mapped onto the
hyperbolic pseudosphere surface (Beltrami trumpet) with negative curvature.
Beltrami trumpet shaped graphene sheets have been predicted to emit Hawking
radiation that is experimentally detectable by a scanning tunnelling
microscope. Here, for the first time we present an analytical algorithm that
allows variational solutions to the Dirac Hamiltonian of graphene
pseudoparticles in BTZ black hole gravitational field by using an approach
based on the formalism of pseudo-Hermitian Hamiltonians within a
discrete-basis-set method. We show that our model not only reproduces the
exact results for the real part of quasinormal mode frequencies of
(2+1)-dimensional spinless BTZ black hole, but also provides analytical
results for the real part of quasinormal modes of spinning BTZ black hole,
and also offers some predictions for the observable effects with a view to
gravity-like phenomena in a curved graphene sheet.
\end{abstract}

\maketitle

After the work of Unruh\cite{Unruh} who first pointed out that quantum fluid
systems can be used to create an analogue of a black hole in the laboratory,
various analogue gravity systems from superfluid $^{3}$He to ultracold
fermions (see e.g. [\onlinecite{Volovik}]) have been proposed to mimic the
Hawking radiation emitted by a black hole. Recently, in a series of papers%
\cite{Iorio11,Iorio12a,Iorio12b,Iorio13,Iorio14} Iorio and Lambiase proposed
a Beltrami trumpet shaped graphene as a tabletop experiment for the quantum
simulation of a black hole, i.e., an analogue black hole to detect the
Hawking-Unruh radiation by a scanning tunneling microscope (STM). The
Beltami trumpet shaped graphene can be thought as a deformed graphene that
have been wrapped up into a hyperbolic pseudosphere with a constant negative
curvature.

(2+1)-dimensional (3D) BTZ black hole model\cite{Banados92} with a negative
cosmological constant $\Lambda =-1/l^{2}$ exhibits similar thermodynamical
properties to those of (3+1)-dimensional Kerr black holes due to having mass
(M), angular momentum (J), and inner and outer horizons. Since it was shown
by Cveti\v{c} and Gibbons\cite{Cvetic12} that the metric of the BTZ black
hole can be mapped onto the Beltrami trumpet, studying the low energy
excitations of Dirac pseudoparticles in a Beltrami trumpet shaped graphene
surface allows one to study quantum properties of the event horizon and
Hawking-like radiation within the BTZ black hole background. The details of
this proposal and its implications are discussed in a comprehensive paper of
Iorio and Lambiase\cite{Iorio14}.

On the gravity side, the first exact solution of quasinormal modes and
associated frequencies for 3D spinless BTZ model were obtained by Cardoso
and Lemos\cite{Cardoso01} and subsequently by Birmingham\cite{Birmingham01},
and to our knowledge no exact or rigorous solutions for quasinormal modes
are available for 3D spinning BTZ black hole. It is well known that, instead
of normal modes, one should deal with quasinormal modes for any dissipative
or non-Hermitian system, as in the case of BTZ black hole, and thus
associated frequencies are complex. For further detailed information on
their interpretation in terms of conformal field theory, we refer to the
paper of Birmingham \textit{et al.}\cite{Birmingham02}, and the references
listed therein.

In particular, the lack of quantum description of the gravitational field,
i.e., quantum gravity, initiated the study of quantum mechanical nature of
black holes, i.e., the quantization of black holes. Within this context,
since Bekenstein's\cite{Bekenstein99} first proposal as an idea of
quantization of the horizon area of nonextremal black holes, many studies
have suggested its use in identifying the role of the real part of
quasinormal frequencies with the certain fundamental quanta of black hole
mass and angular momentum (see Ref.~[\onlinecite{Dreyer03}] and the
references therein). Among them, the first major conceptual contribution was
made by Hod\cite{Hod98} who showed that the real part of the quasinormal
frequencies should be related to this quantized black hole surface area,
leading to a strong evidence in favor of uniformly spaced spectrum of
quantum black holes. In this regard, all the studies made so far in this
field are concerned with the spinless black holes, since as outlined above
achieving quasinormal mode spectrum in the presence of spin is still a
challenging problem.

In this paper, we propose an experimentally realizable model to study the
emergence of gravity-like phenomena in curved graphene sheet. First, we
solve the non-Hermiticity problem of a Hamiltonian for Dirac pseudoparticles
in the presence of BTZ gravitational field by well-known pseudo-Hermitian
quantum mechanical tools\cite{Gorbatenko10, Gorbatenko11}, and then
investigate its variational solutions by an unconventional
discrete-basis-set method\cite{Goldman81,Doran05}. Besides, in the context
of BTZ black hole solutions, we obtained rigorous analytical results for the
real part of quasinormal mode frequencies for a spinning BTZ black hole that
cover those for spinless one, $J=0$ for the lowest mode number, namely $n=0$%
, which had been previously found in the literature\cite%
{Cardoso01,Birmingham01, Birmingham04,Myung12,Becar14}. Moreover, for a
particular choice of black hole parameters, our solutions not only reduce to
static spinless case, but also present analytical expressions for vacuum,
anti de Sitter and extremal cases.

The BTZ black hole solutions of the 3D Einstein gravity are given by the
axially symmetric metric\cite{Banados92} 
\begin{equation}
ds^{2}=-\Delta (r)dt^{2}+\Delta ^{-1}(r)dr^{2}+r^{2}\left( d\phi +\Delta
^{\phi }(r)dt\right) ^{2}  \label{1}
\end{equation}%
with the lapse function $\Delta (r)$ and the angular shift $\Delta ^{\phi
}(r)$ 
\[
\Delta (r)=\frac{r^{2}}{l^{2}}+\frac{J^{2}}{4r^{2}}-M\quad ,\quad \Delta
^{\phi }(r)=-\frac{J}{2r^{2}} 
\]%
where $M$ and $J$ are the mass and angular momentum (spin) of the BTZ black
hole, respectively, and $l$ is the cosmological length that is related to
the negative cosmological constant $\Lambda $ by $l=\sqrt{-\Lambda }$. Eq.(%
\ref{1}) has two coordinate singularities corresponding to the inner ($r_{-}$%
) and outer ($r_{+}$, or event horizon of the black hole) horizons of the
black hole, respectively, and they are given as 
\begin{equation}
r_{\mp }=l\left( \frac{M}{2}\left\{ 1\mp \left[ 1-\left( \frac{J}{Ml}\right)
^{2}\right] ^{1/2}\right\} \right) ^{1/2}.  \label{2}
\end{equation}%
The event horizon of the black hole exist only under the conditions, $M>0$, $%
|J|\leq Ml$. Since the static metric can be conformally rescaled up to a
conformal factor $\Delta (r)$, it is easy to show that its scaled ultra
static part is conformal to the axisymmetric optical Zermelo metric\cite%
{Cvetic12}, and it can be mapped onto Beltrami trumpet surface with constant
negative curvature. This conformal relation allows one to study the effects
of horizons on low energy excitations of massless Dirac fermions on a curved
graphene sheet by considering the associated massless Dirac equation
subjected to externally applied electric field and gravito-magnetic field in
the BTZ metric background. The resultant Dirac equation in the optical BTZ
black hole background can be written\cite{Cvetic12} as%
\begin{eqnarray}
&&\left[ \sigma _{1}\left( \Delta \frac{\partial }{\partial r}-\frac{M}{2r}+%
\frac{J^{2}}{4r^{3}}\right) +\sigma _{3}\frac{im}{r}\Delta ^{1/2}\right. 
\nonumber \\
&&\qquad \qquad \quad \left. -\sigma _{2}\left( -\epsilon+m\frac{J}{2r^{2}}%
\right) -\frac{J}{4r^{2}}\Delta ^{1/2}\right]\psi (r) =0,  \nonumber \\
&&  \label{3}
\end{eqnarray}%
where $\sigma _{i}$ are the Pauli spin matrices, and axial symmetry in
stationary picture is taken into account by choosing the wave function $\Psi
=\psi (r)e^{-i\epsilon t+im\phi }$ with $m$ is the azimutal angular quantum
number.

In general, the associated Hamiltonian describing the dynamics of spin $1/2$
particles in a gravitational background such as in Eq.(\ref{3}) has no
hermiticity. Fortunately, it is shown very recently that, the hermiticity
problem of Hamiltonians corresponding to a spin $1/2$ particle in an axially
symmetric stationary gravitational background can be handled by using the
pseudo-Hermitian quantum mechanical tools\cite{Gorbatenko10, Gorbatenko11}.
If one finds an invertible operator $\rho $ satisfying the Parker weight
operator relationship $\rho =\eta ^{\dagger }\eta $, then it is easy to find
a Hermitian Hamiltonian such as $\mathcal{H}_{\eta }=\eta \mathcal{H}\eta
^{-1}=\mathcal{H}_{\eta }^{\dagger }$ whose spectrum coincides with those of 
$\mathcal{H}$. The Hermitian Hamiltonian can be written in terms of metric
components $g_{ab}$ and the determinant of the metric $g$ as follows 
\begin{eqnarray}
\mathcal{H}_{\eta } &=&\mathcal{H}_{0}-\frac{i}{4(-g^{00})}\widetilde{\gamma 
}^{0}\widetilde{\gamma }^{k}\left[ \frac{\partial \ln (-g)}{\partial x^{k}}+%
\frac{\partial \ln (-g^{00})}{\partial x^{k}}\right]  \nonumber \\
&&\qquad+\frac{i}{4}\left[ \frac{\partial \ln (-g)}{\partial t}+\frac{%
\partial \ln (-g^{00})}{\partial t}\right]  \label{4}
\end{eqnarray}%
with 
\[
\mathcal{H}_{0}=\frac{i}{\left( -g^{00}\right) }\widetilde{\gamma }^{0}%
\widetilde{\gamma }^{k}\frac{\partial }{\partial x^{k}}-i\widetilde{\Phi }%
_{0}+\frac{i}{\left( -g^{00}\right) }\widetilde{\gamma }^{0}\widetilde{%
\gamma }^{k}\widetilde{\Phi }_{k} 
\]%
where 
\begin{eqnarray}
\widetilde{\gamma }^{\alpha } &=&\widetilde{H}_{\underline{\beta }}^{\alpha
}\gamma ^{\underline{\beta }}  \label{5} \\
\widetilde{\Phi }_{\alpha } &=&-\frac{1}{4}\widetilde{H}_{\mu }^{\underline{%
\epsilon }}\widetilde{H}_{\nu \underline{\epsilon };\alpha }\widetilde{S}%
^{\mu \nu }  \label{6}
\end{eqnarray}%
are curved space gamma matrices and transformed bispinor connectivity,
respectively, together with $S^{\mu \nu }=\frac{1}{2}\left( \widetilde{%
\gamma }^{\mu }\widetilde{\gamma }^{\nu }-\widetilde{\gamma }^{\nu }%
\widetilde{\gamma }^{\mu }\right) $. In Eqs.(\ref{4}-\ref{6}), $\widetilde{%
\gamma }^{\alpha }$ are determined in terms of Dirac matrices with local
indices $\gamma ^{\underline{\alpha }}$ through the transformed tetrad
vectors $\widetilde{H}_{\underline{\beta }}^{\alpha }$ in Schwinger gauge,
and they are also related to the Dirac matrices with global indices $\gamma
^{\alpha }$ through Schwinger functions of the corresponding metric by $%
\gamma ^{\alpha }=H_{\underline{\beta }}^{\alpha }\gamma ^{\underline{\beta }%
}$. Here the indices take values from $0$ to $2$, and semicolon denotes the
covariant derivative which is written as 
\begin{equation}
H_{\nu \underline{\epsilon };\alpha }=\frac{\partial H_{\nu \underline{%
\epsilon }}}{\partial x^{\alpha }}-\Gamma _{\nu \alpha }^{\lambda
}H_{\lambda \underline{\epsilon }}  \nonumber
\end{equation}%
where $\Gamma _{\nu \underline{\epsilon }}^{\lambda }$ denotes the
connection coefficients and $H_{\lambda \underline{\epsilon }}=g_{\lambda
\mu }H_{\underline{\epsilon }}^{\mu }$. By using the components of the
Zermelo metric conformally related to Eq.(\ref{1}), one finds the components
of the transformed bispinor connectivity given by Eq.(\ref{6}) as 
\begin{eqnarray}
\widetilde{\Phi }_{0} &=&\frac{J^{2}}{16r^{3}}\left( \gamma ^{\underline{2}%
}\gamma ^{\underline{0}}-\gamma ^{\underline{0}}\gamma ^{\underline{2}%
}\right) +\frac{J\sqrt{\Delta }}{16r}\frac{\Delta ^{\prime }}{\Delta }\left(
\gamma ^{\underline{1}}\gamma ^{\underline{2}}-\gamma ^{\underline{2}}\gamma
^{\underline{1}}\right)  \nonumber \\
\widetilde{\Phi }_{1} &=&\frac{J}{8r^{2}\sqrt{\Delta }}\left( \gamma ^{%
\underline{0}}\gamma ^{\underline{1}}-\gamma ^{\underline{1}}\gamma ^{%
\underline{0}}\right)  \nonumber \\
\widetilde{\Phi }_{2} &=&\frac{J}{8r}\left( \gamma ^{\underline{0}}\gamma ^{%
\underline{2}}-\gamma ^{\underline{2}\gamma ^{\underline{0}}}\right) 
\nonumber \\
&&+\left( \frac{\sqrt{\Delta }}{4}-\frac{r\Delta ^{\prime }\sqrt{\Delta }}{%
8\Delta }\right) \left( \gamma ^{\underline{1}}\gamma ^{\underline{2}%
}-\gamma ^{\underline{2}}\gamma ^{\underline{1}}\right) .  \label{7}
\end{eqnarray}%
Therefore, by replacing Eq.(\ref{7}) back into Eq.(\ref{4}),\ \ the
Hermitian Dirac Hamiltonian corresponding to the BTZ black hole background
can be found as 
\begin{equation}
\mathcal{H}_{\eta }=i\sigma _{2}\left( \Delta \frac{\partial }{\partial r}+%
\frac{\Delta ^{\prime }}{2}\right) +m\frac{J}{2r^{2}}-\sigma _{1}m\frac{%
\sqrt{\Delta }}{r}+\sigma _{3}\frac{J\sqrt{\Delta }}{4r^{2}},  \label{8}
\end{equation}%
which yields the energy eigenvalues of \ Dirac pseudoparticles on a Beltrami
trumpet shaped graphene. In Eqs.(\ref{7}) and (\ref{8}), the prime shows the
derivative of the lapse function with respect to $r$ and within the
framework of 3D relativistic field theory, the Hamiltonian was written in
units of $\hbar c$, where $c$ is the velocity of light that will be replaced
by the Fermi velocity $v_{F}$ in graphene.

It is well-known that the conventional Rayleigh-Ritz variational method can
not be directly applied to the Dirac equation, due to the absence of \ both
upper and lower bounds of the associated Dirac Hamiltonian. The Dirac
Hamiltonians are not bounded from below so that spurious roots may appear
due to variational instability. This is the second obstacle that one faces
when dealing with Dirac Hamiltonians. To overcome this difficulty, Drake and
Goldman \cite{Goldman81} have firstly proposed a discrete-basis-set method
to eliminate the spurious roots. This variational procedure is based on
defining a two component radial spinor as a trial function of the form 
\begin{equation}
|\Phi \rangle =g(r)\left( 
\begin{array}{c}
a \\ 
b%
\end{array}%
\right)   \label{9}
\end{equation}%
where $a$ and $b$ are variational parameters, and $g(r)$ is an arbitrary
continuous function satisfying the conditions $\int_{0}^{\infty }g^{2}(r)dr=1
$ and $\lim_{r\rightarrow \infty }g(r)=0$. We now introduce an \textit{ansatz%
} for the normalized basis function to get an approximation to the
ground-state of the system as 
\[
g(r)=2(\sqrt{M}l)^{-3/2}re^{-r/{\sqrt{M}l}}
\]%
which satisfies the necessary constraints to avoid the spurious roots. Eq.(%
\ref{9}) yields the matrix representation of the Hamiltonian in Eq.(\ref{8}%
). Thus, the associated eigenvalue equation stands for the condition $(%
\mathcal{H}_{\eta }-\epsilon )|\Phi \rangle =0$ where $\epsilon $ is the
energy relative to $\hbar c$ and it is real. With this condition, and the
variation with respect to $a$ and $b$, the resulting set of linear equations
can be self-consistently solved if and only if the corresponding
two-dimensional characteristic determinant is set to be equal to zero.
Finally, we find eigenvalues of $\mathcal{H}_{\eta }$ in the form 
\begin{equation}
\epsilon =\frac{1}{2}m\mathrm{J\ P}\pm \frac{1}{4}\sqrt{\mathrm{J}^{2}%
\mathrm{Q}^{2}+16m^{2}\mathrm{R}^{2}}  \label{10}
\end{equation}%
where $\mathrm{P}$, $\mathrm{Q}$ and $\mathrm{R}$ are defined as integrals 
\begin{eqnarray}
\mathrm{P} &=&\int_{r_{+}}^{\infty }\frac{g^{2}(r)}{r^{2}}dr  \nonumber \\
\mathrm{Q} &=&\int_{r_{+}}^{\infty }\frac{g^{2}(r)}{r^{2}}\sqrt{\Delta (r)}dr
\nonumber
\end{eqnarray}%
and 
\[
\mathrm{R}=\int_{r_{+}}^{\infty }\frac{g^{2}(r)}{r}\sqrt{\Delta (r)}dr,
\]%
in terms of cutoff function given by Eq.(\ref{2}). In Eq.(\ref{10}), upon
the choice of $M$ and $J$ values, following four special cases of the BTZ
black hole can easily be found:

(i) For the vacuum state ($M=0$, $J=0$), Eq.(\ref{10}) reduces to 
\begin{equation}
\epsilon=\pm \frac{m}{l}  \label{11}
\end{equation}

(ii) For the static black hole ($M\neq 0$, $J=0$) Eq.(\ref{10}) gives 
\begin{equation}
\epsilon=\pm \frac{m}{l}\frac{1}{2}K_{2}(2)  \label{12}
\end{equation}%
where $K_{2}$ is the modified Bessel function of the second kind.

Here, it should be mentioned that energy eigenvalues Eq.(\ref{11}-\ref{12})
for the spinless BTZ black hole are similar to those found for the real part
of quasinormal frequencies in Refs.[%
\onlinecite{Cardoso01,Birmingham01,Birmingham04,Myung12,Becar14}].

(iii) For the case $M=-1$, $J=0$ which is recognized as anti-de Sitter (AdS)
spacetime, Eq.~(\ref{11}) gives 
\begin{equation}
\epsilon=\pm \frac{m}{l}\frac{1}{3}\left[ H_{2}(2)-3\pi Y_{2}(2)-4\right]
\label{13}
\end{equation}%
where $H_{2}$ and $Y_{2}$ are Struve functions, and second kind of Bessel
functions, respectively.

(iv) For the extremal BTZ black hole ($M\neq 0$, $J=Ml$) Eq.~(\ref{10})
becomes 
\begin{equation}
\epsilon=\frac{1}{l}\left[ m\pm \sqrt{M\Theta _{1}^{2}+m^{2}\Theta _{2}^{2}}%
\right]  \label{14}
\end{equation}%
where we have defined $\Theta _{1}=\left[ (1+\sqrt{2})\exp (-\sqrt{2})+2Ei(-%
\sqrt{2})\right] /4$ and $\Theta _{2}=(5+3\sqrt{2})\exp (-2\sqrt{2})$ in
terms of exponential integral $Ei$.

In each of the first three cases, Eqs.(\ref{11}-\ref{13}) leads to a
sequence of different ground-state energies depending on the azimuthal
quantum number. They are equally spaced below and above the Dirac point, $%
m=0 $. In the latter case, i.e., extreme black hole case, in Eq.(\ref{14})
there appears a gap which is equal to $2\sqrt{M}\Theta _{1}/l$.

In the context of graphene, by using the graphene units, Eq.(\ref{10}) can
be rewritten in terms of external electric and magnetic fields together with
a gap term as follows 
\begin{equation}
E=em\Phi _{m}(\bar{J})\pm \sqrt{\Delta ^{2}(\bar{J})+m^{2}(\hbar \omega
_{B})^{2}},  \label{15}
\end{equation}%
where 
\begin{eqnarray}
\Phi (\bar{J}) &=&\frac{4\bar{J}}{\delta }e^{-2\bar{r}_{+}}(\text{V}), 
\nonumber \\
\Delta (\bar{J}) &=&\frac{4\bar{J}}{\delta ^{2}}\bar{\mathrm{Q}}(\bar{r}_{+},\bar{J%
})\,(\text{meV}),  \label{16}
\end{eqnarray}%
and frequency $\omega _{B}=\sqrt{2}v_{F}/l_{B}$ is the cyclotron frequency
in the relativistic case such that energy has $\sqrt{B}$ magnetic field
dependence with 
\begin{equation}
B=\frac{4B_{0}}{\delta ^{2}}\bar{\mathrm{R}}^{2}(r_{+},\bar{J})(\text{T})
\label{17}
\end{equation}%
where $B_{0}=3,26\times 10^{4}$T for graphene. In Eqs.(\ref{16}) and (\ref%
{17}) , while $\delta $ denotes the ratio of cosmological length to the
lattice parameter of graphene and $\bar{J}=J/\sqrt{M}l$ is the dimensionless
spin that changes in the interval between 0 and 1, $\bar{r}_{+}=r_{+}/\sqrt{M%
}l=\left[ 1+\left( 1-\bar{J}^{2}\right) ^{1/2}\right] ^{1/2}/\sqrt{2}$ is
the dimensionless event horizon, the dimensionless integrals $\bar{\mathrm{Q}}$
and $\bar{\mathrm{R}}$ are given by 
\begin{eqnarray*}
\bar{\mathrm{Q}}(\bar{r}_{+},\bar{J}) &=&\int_{\bar{r}_{+}}^{\infty }d\bar{r}e^{-2%
\bar{r}}\sqrt{\bar{r}^{2}-1+\frac{\bar{J}^{2}}{4\bar{r}^{2}}} \\
\bar{\mathrm{R}}(\bar{r}_{+},\bar{J}) &=&\int_{\bar{r}_{+}}^{\infty }\bar{r}d\bar{r%
}e^{-2\bar{r}}\sqrt{\bar{r}^{2}-1+\frac{\bar{J}^{2}}{4\bar{r}^{2}}}
\end{eqnarray*}%
respectively. In writing the gap term in Eq.(\ref{16}) , we use the \textit{%
ansatz} of Iorio and Lambiase\cite{Iorio14}, i.e., $\sqrt{M}=1/\bar{l}$, to
be able to write the relevant BTZ quantities in terms of graphene
parameters. Therefore, instead of making a Beltrami trumpet shaped graphene
to observe Hawking radiation by a STM, we model it with position dependent
electric and magnetic fields together with a position dependent mass-like
term in a flat graphene sheet.

\begin{figure}[tbp]
\includegraphics[scale=0.8]{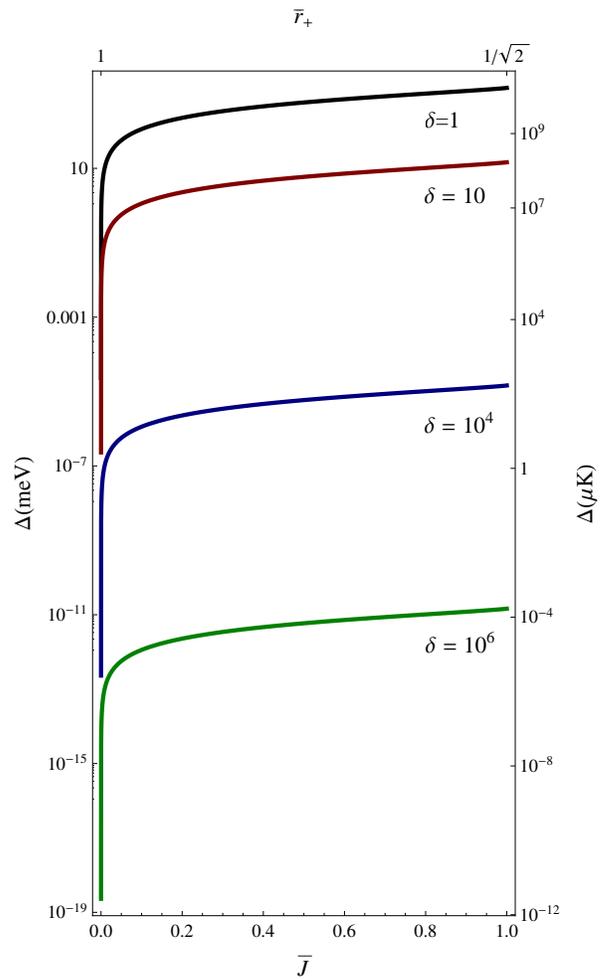}
\caption{(Color online) Dependence of the half band gap as a function of the
black hole spin for four different values of $\protect\delta$. }
\label{FIG1}
\end{figure}

In FIG.~\ref{FIG1}, we plot the dimensionless spin dependence of the
resulting mass-like term, i.e., half gap for different values of $\bar{l}$
which are chosen as to be compatible with those in Ref.[\onlinecite{Iorio14}%
]. It is easy to see from the figure that the resulting gap can be
controlled via dimensionless spin. Note that the cut-off function appearing
in integrals of Eqs.(\ref{16}) and (\ref{17}) is an implicit function of
dimensionless spin $\bar{J}$ whose domain lies between 0 and 1. This implies
that $\bar{J}$ is a function of $\bar{r}_{+}$ whose range should lie between 
$1$ and $1/\sqrt{2}$.

As a conclusion, in this letter, we have, for the first time, obtained the
quasinormal modes of both spinning and spinless BTZ black holes by using
pseudo-Hermitian quantum mechanical tools within the discrete-basis-set
variational method, and proposed an experimental table-top setup allowing to
construct an analogy between gravity and condensed matter physics.
Therefore, our findings are two-fold. First, the analytical results we
obtained for the spinning BTZ do recover the well-known results for the
spinless one\cite{Cardoso01,Birmingham01, Birmingham04,Myung12,Becar14},
and thus they are consistent with those found in the literature. We show
that spinless BTZ black hole solutions have equally spaced eigenvalues,
whereas the spectrum of rotating BTZ black hole is again discrete, but looks
Dirac-like which includes unequally spaced levels. Additionally, due to the
conformal relation between the metric of the BTZ black hole and the low
energy excitations of Dirac pseudo-particles in a Beltrami shaped graphene,
our method developed here allows us to propose an experimental scheme to
observe Hawking-like effect in a graphene sheet in the presence of uniform
magnetic and electric fields together with a gap opening term. We showed
theoretically that different electrical and magnetic field profiles induce
different black hole regimes, and they can be used experimentally to imitate  quantum properties of a 3D BTZ black hole in a lab.

\acknowledgments

This work is supported by the Scientific and Technological Research Council
of Turkey (T\"{U}B\.{I}TAK) grant number 113F103. The authors thank Prof.Dr.
S. Ata\u{g} for valuable discussions and comments.


\end{document}